\documentclass[fleqn,usenatbib]{mnras}
\pdfoutput=1
\usepackage{mathptmx}
\usepackage{rotating}
\hypersetup{draft}
\usepackage[T1]{fontenc}
\usepackage[english]{babel}
\usepackage{ae,aecompl}
\usepackage{float}
\usepackage{graphicx}   
\usepackage{amsmath}	
\usepackage{amssymb}	
\usepackage{latexsym}
\usepackage{times}
\usepackage{subfig}
\usepackage{multirow}
\usepackage{array}
\usepackage{csquotes}
\usepackage{tikz}
\newcommand{\be}{\begin{equation}}
\newcommand{\ee}{\end{equation}}
\title{Revealing the Unusual Structure of the KAT-7-Discovered Giant Radio Galaxy J0133$-$1302}
\author[N. Mhlahlo et al.]{
N. Mhlahlo,$^{1}$\thanks{E-mail: nceba.mhlahlo@wits.ac.za} M. Jamrozy,$^{2}$\thanks{E-mail: jamrozy@oa.uj.edu.pl}
\\
$^{1}$School of Physics, University of the Witwatersrand, Private Bag 3, 2050-Johannesburg, South Africa   \\
$^{2}$Astronomical Observatory, Jagiellonian University, ul. Orla 171, 30-244 Krakow, Poland 
}
\date{Accepted XXX. Received YYY; in original form ZZZ}
\pubyear{2019}

\begin{document}
\label{firstpage}
\pagerange{\pageref{firstpage}--\pageref{lastpage}}
\maketitle
\begin{abstract}
We present a new study of the 1.7 Mpc KAT-7-discovered giant radio galaxy, J0133$-$1302, which was carried out using GMRT data at 323 and 608 MHz. This source is located at RA $01^h33^m13^s$ and Dec $-13^{\circ}03^\prime00^{\prime\prime}$ and has a photometric redshift of $\sim$0.3.
We discovered unusual morphological properties of the source which include lobes that are exceptionally asymmetric, where the upper lobe is much further from the core when compared to the lower lobe, and a complex structure of the upper lobe. This complex structure of the upper lobe hints at the presence of another source, in close proximity to the edge of the lobe, which resembles a bent-double, or distorted bent tail (DBT) radio galaxy.  
Both the upper lobe and the lower lobe have a steep spectrum, and the synchrotron age of the lower lobe should be less than about 44 Myr. 
The core has an inverted spectrum, and our results suggest that the parent galaxy in J0133$-$1302 is starting a new jet activity. 
Our spectral analysis indicates that this source could be a GigaHertz Peaked Spectrum (GPS) radio galaxy.  
\end{abstract}
%
\section{Introduction}
Radio galaxies (RGs) are extragalactic radio sources that exhibit extended radio emission on either side of the active galactic nucleus (AGN). These sources are characterised by the presence of jets, which are usually two collimated outflows from the AGN, and the radio lobes where the jets terminate (\citealt{Mil80, Gio05, Lis16}; see also the recent comprehensive review by \citealt{Har20}). The jets carry plasma away from the nucleus to the opposite ends of the galaxy, additionally leading to flows of momentum, energy and magnetic flux \citep{Beg84, Lai05}.
The radio lobes, and the jets when they are present, can track the largest sources among radio galaxies known as giant radio galaxies (GRGs). GRGs represent the most energetic single-galaxy structures in the universe, with jets and lobes extending on $\sim$Mpc scales \citep{Ish99, Lar01}. These sources belong mostly to the Fanaroff-Riley type II (FRII; \citealt{Fan74}) radio morphology \citep{Dab20}, having high radio luminosities when compared to the Fanaroff-Riley type I (FRI) radio sources (\citealt{Lai05}; but see the discussion by \citealt{Ming19} about the FRI/II luminosity break). They have been observed at multifrequencies using a variety of radio instruments \citep{Ish99, Lar01a, Mac01, Ish02, Kro04, Sar05, Kon08, Jam08, Mac09, Mal13, Ming19, Mac20, Cot20, Dab20}.   
The GRGs are essential for the investigation of the astronomical questions about the evolution of radio sources, the acceleration of cosmic rays, the density of the intergalactic medium at different redshifts, and the nature of the AGN activity. The largest known GRG is J1420-0545 with a projected linear size of 4.69 Mpc (\citealt{Mac08}; but also see \citealt{Her17} for a potentially larger GRG). 
To date it is still not clear what leads to the formation of GRGs. Two fundamental scenarios have been postulated to describe their exceptional sizes: firstly, the lobes could be fed by very powerful central engines which supply the jets with enough energy to bore their way through the ambient medium \citep{Gop89}, and secondly, GRGs could be normal radio sources evolving in very low-density environments that offer small resistance to the evolution of the jets {\citep{Lar00,Sub08}. The first scenario requires the presence of prominent cores, due to strong nuclear activity, which are not always observed. \cite{Ish99} investigated this possibility and found an inverse correlation between the degree of core prominence and total radio luminosity, which indicated that the giant radio objects have similar core strengths to those of smaller objects of similar total luminosity.  \\ A number of case and comparison studies of the properties of GRGs and the ambient intergalactic medium \citep{Lar00, Sub08, Saf09, Har20} seem to point to the second scenario. However, a study by \cite{Kom09} found that for a sample of GRGs, the asymmetries in the radio morphology of a several GRGs appeared to be related to the density anisotropy in the medium.
If this observation is indeed correct, it could imply that a low environmental density cannot be the only origin of GRGs. \\
Most of the normal-size RGs ($<$ 1 Mpc), and GRGs, have their lobes continuosly supplied with energy from the AGN via the jets. 
However, a number of studies have shown that there are sources where the nuclear activity has stopped, leading to the interruption of the jet production mechanism. This cuts off the energy flow from the AGN, and results in a gradual fading of structures due to adiabatic and radiative losses of relativistic electrons (e.g. J0349+7511 in Abell 449; \citealt{Hun16}). The resulting diffuse sources are sometimes referred to as relics, AGN relics, or loosely as dying sources \citep{Cor87, Ren97, Jam04, Par07, Mur11, Tam15, Hur15, Bri16, Shu17, Mah18, Ran20, Shab20, Qui21}.   
They are characterised by a steep spectrum with a high-energy cut-off or turnover at some characteristic break frequency.  \\
Curiously, some otherwise dying sources have shown hints of recurrent or restarting AGN activity.  It has been established beyond doubt that jet forming activity in these sources is episodic and not continuous during the lifetime of the source.
These sources are called double-double RGs (DDRGs), and their morphology carries information about their interrupted jet activity and evolutionary history. A characteristic feature is a pair of double radio sources emanating from the AGN, and this is thought to occur when a new epoch of jet activity, after a quiescent or \enquote{off} phase, takes place in a RG with older lobes still visible from the activity of the previous epoch (\citealt{Sch00, Mar16, Hun16, Ming19, Shab20}; and for the recent reviews on DDRGs see \citealt{Sai09,Kuz17,Mah19}). \\
Since the lobes of extended radio sources can store the energy supplied by the jets for longer than the duration of the \enquote{off} phase, radio galaxies are able to preserve information on the past activity and history of the AGN \citep{Jam12}.
%
\begin{figure*}
\vspace{0.7cm}
\begin{tabular}{cc}
\centering
\includegraphics[scale=0.5]{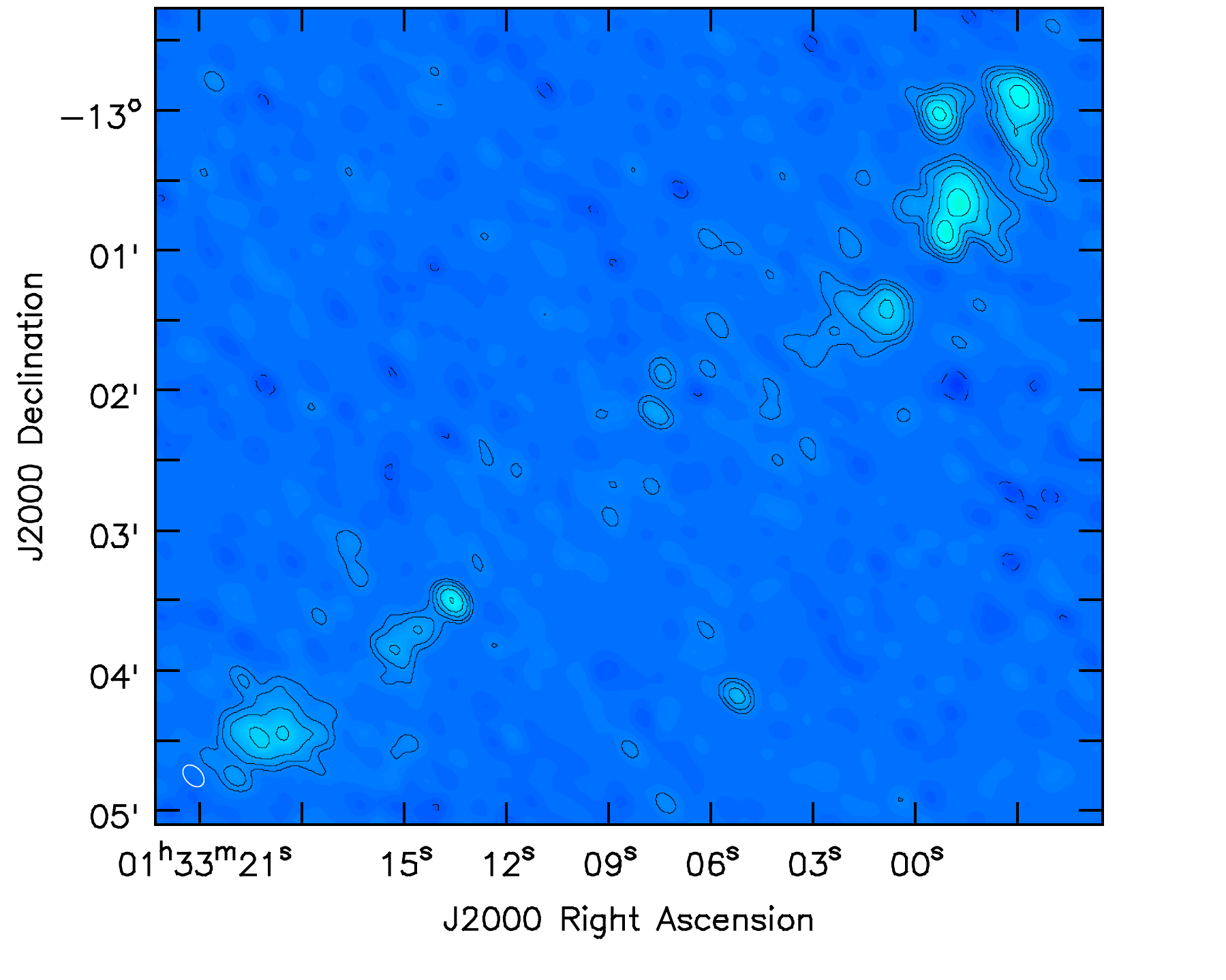}\vspace{-0.0cm} &
\includegraphics[scale=0.5]{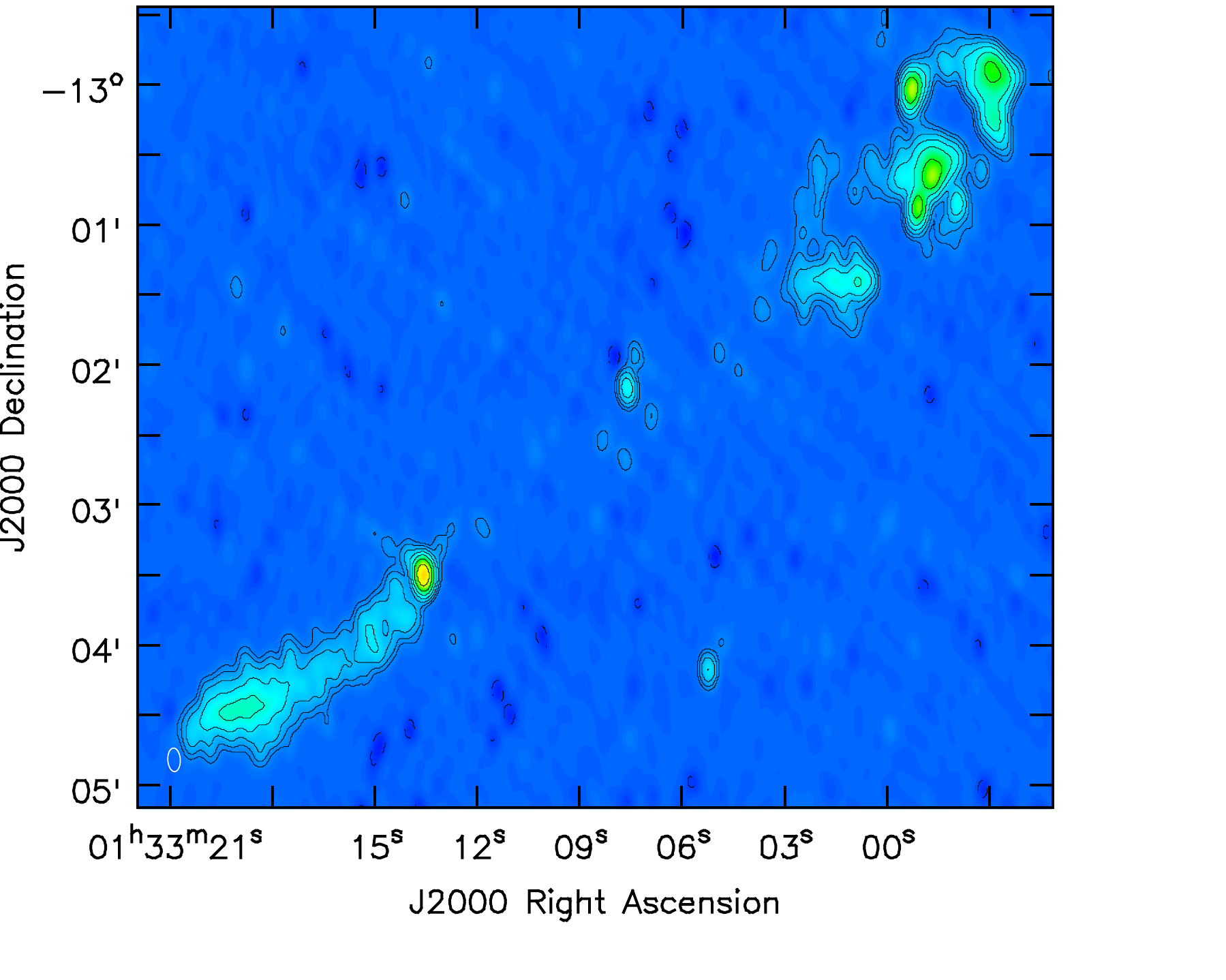}\vspace{-0.0cm} \vspace{0.6cm} \\ 
\includegraphics[scale=0.5]{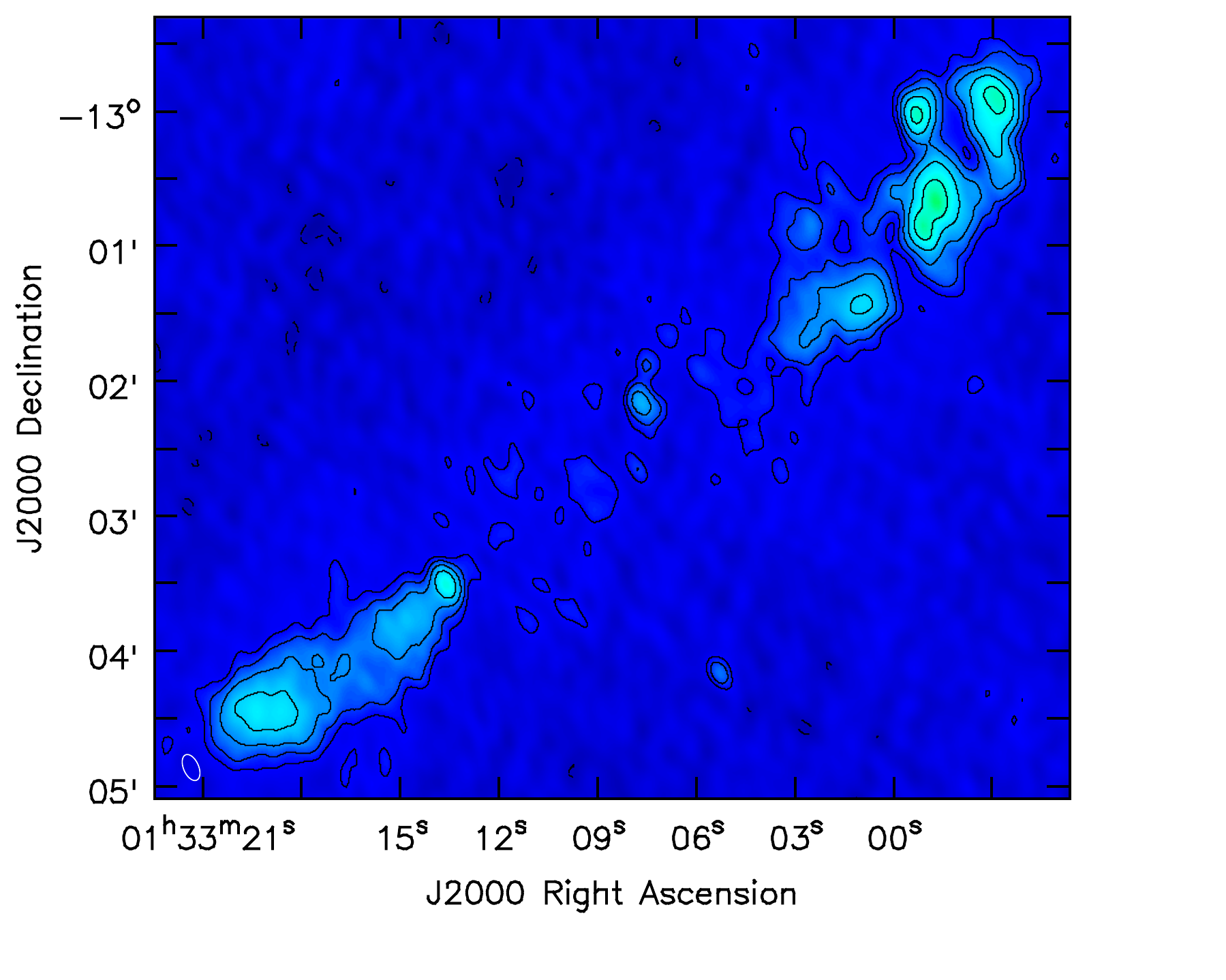} & 
\includegraphics[scale=0.5]{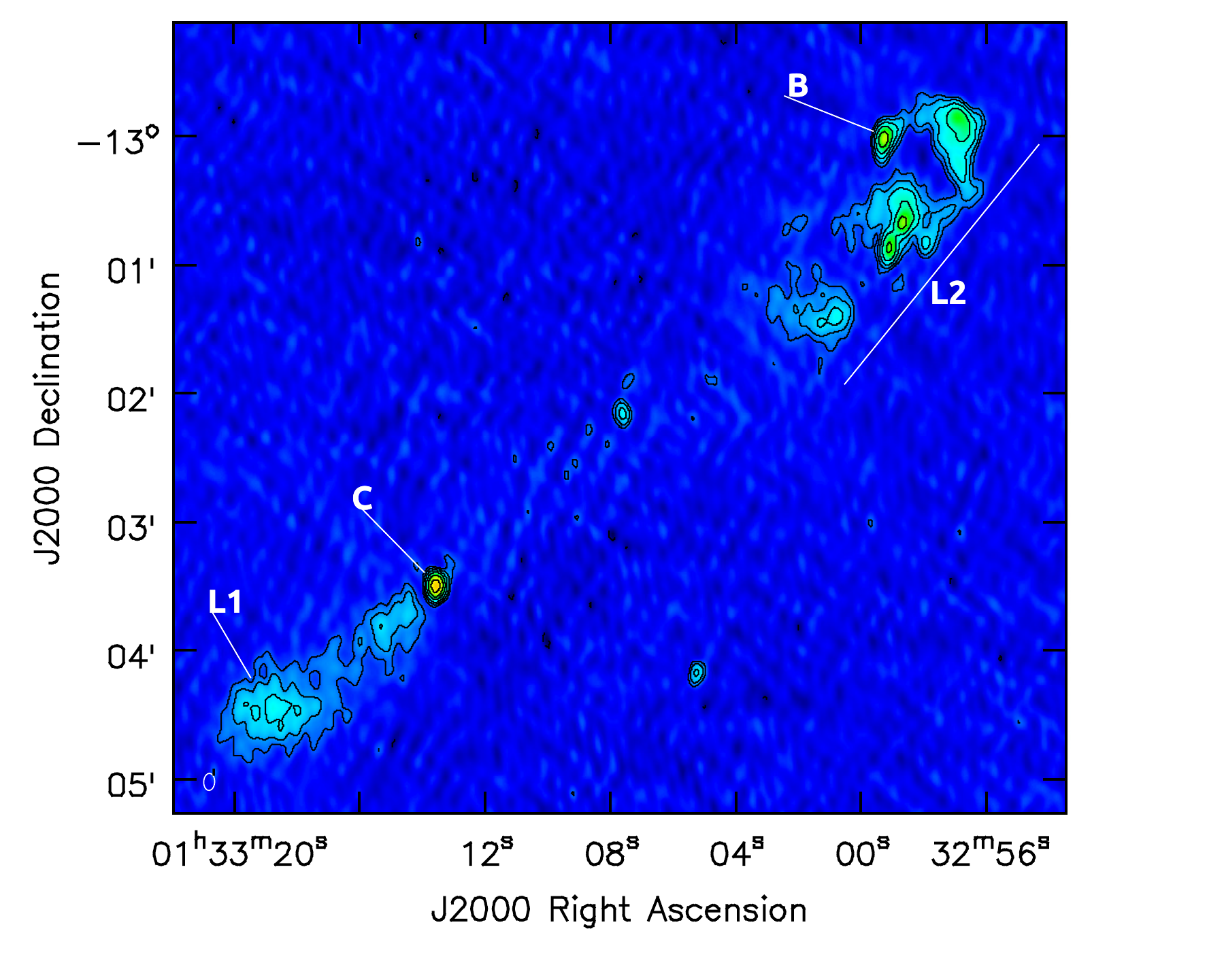}
\end{tabular}
\vspace{-0.6cm}
\caption{Top left: AIPS radio image of the GRG at 323 MHz. The restoring beam of the image is 11$^{\prime\prime}$ $\times$7 $^{\prime\prime}$ (PA=43 deg), and the rms noise level in the image plane is 0.1 mJy/beam. Top right: AIPS radio image at 608 MHz. The restoring beam is 10$^{\prime\prime}$ $\times$5 $^{\prime\prime}$ (PA=3.3 deg), and the rms noise is 0.03 mJy/beam.  \\ 
Bottom: SPAM radio images of the GRG at 323 MHz (left) and 608 MHz (right) are shown. The restoring beam of the image at 323 MHz is 12$^{\prime\prime}$ $\times$ 7$^{\prime\prime}$ (PA=24 deg), and that of the image at 608 MHz is 8$^{\prime\prime}$ $\times$ 5$^{\prime\prime}$ (PA=-2 deg). The noise level in the image plane is 0.1 mJy/beam (left) and 0.08 mJy/beam (right). In all the images radio contours start at $3\times rms$ and then scale by a factor of 2. The restoring beam size is shown at the bottom-left hand corner. L1 represents the south-eastern radio lobe and L2 the north-western lobe; C stands for the radio core of the GRG and B for a background/foreground source or a hotspot.}
\label{fig:aips-spam-maps}
\end{figure*}
\section{GRG J0133$-$1302}
GRG J0133$-$1302 was discovered in the field of the cluster of galaxies ACO209 at a frequency of 1.83 GHz \citep{Col16}, using the decomissioned 7-dish Karoo Array Telescope (KAT-7), located near the Square Kilometer Array (SKA) core site in the Northern Cape Karoo, South Africa \citep{Fol16}. This source is one of the eight target clusters which were observed with the KAT-7. It is located at RA $01^h33^m13^s$ and Dec $-13^{\circ}03^\prime00^{\prime\prime}$ (J2000.0) and has a redshift of $\sim$0.3 \citep{Col16}. This is, however, a photometric redshift which might have a large error.  \\
The KAT-7 radio telescope array detected extended emission in the form of two symmetric lobes.  A cross-correlation of the KAT-7, NRAO VLA Sky Survey (NVSS: \citealt{Con98}), infra-red (IR), optical, and X-ray sources in the field of GRG J0133$-$1302 revealed the KAT-7 extended source encompassing NVSS sources. The NVSS sources are S1 (southern-east SE lobe), S2 (core), S3 and S4 (northern-west NW lobe), according to the nomenclature by \cite{Col16}, suggesting there was an astrophysical connection between the lobes and the central core of an extended radio galaxy. This means the NVSS image partially resolves the elongated KAT-7 radio structure into four distinct sources (see Fig.\,1 and 2 in \citealt{Col16}). The NVSS core is associated with the X-ray source 1RXSJ013313.8-13031 in the ROSAT all-sky survey (RASS) having a count rate of 0.04 cts/s (0.1-2.4 keV band). \\
At the position of the radio core, or host galaxy, there is also a strongly detected WISE source J013313.50-130330.5. 
The infrared magnitudes of the host galaxy from the Wide-field Infrared Survey Explorer (WISE; \citealt{Wri10}) and from Two Micron All-Sky Survey (2MASS; \citealt{Skr06}) are as folows: W1(3.4$\mu m$)=14.1, W2(4.6$\mu m$)=13.2, W3(11.6$\mu m$)=10.7, W4(22$\mu m$)=8.5 and K$_{s}$=15.1, H=15.4, J=16.7. The optical magnitudes taken by the UK Schmidt Telescope and outlined by the SuperCOSMOS Sky Surveys \citep{Ham01} are I=17.5,  R=18.1 and B(J)=19.6. In addition, the Panoramic Survey Telescope \& Rapid Response System (Pan-STARRS; \citealt{Fle20}) five bands magnitudes are y=18.5, z=18.6,  i=19.1, r=19.5 and g=20.2. In two ultraviolet bands, near-UV (NUV) and far-UV (FUV), the Galaxy Evolution Explorer (GALEX; \citealt{Mar05}) imaged the host galaxy, and their magnitudes are NUV=21.4, FUV=22.4.  \\
\cite{Wri10} in their Fig.\,12 show a colour-colour diagram with regions where the different classes of the WISE-detected sources are located. The host of J0133$-$1302 has the following WISE colors (W1$-$W2)$=$0.9 and (W2$-$W3)$=$2.5 and therefore belongs to the quasar class.  According to \cite{Izo14}, galaxies with (W1$-$W2) colours greater than 1 magnitude are mainly luminous galaxies with high-excitation HII regions. In addition, \cite{Col16} have reported that the host galaxy is associated with the ROSAT X-ray source 1RXSJ013313.8-13031. All of this indicates that this host is a powerful AGN. Taking into account colours in the other bands indicates that this object has red continua with $J-K=1.6$ and $g-r=0.7$.  \cite{Urr09} and \cite{You08} classified objects possessing colours $J-K > 1.3$ and $g-r> 0.5$ as red quasars. This class of quasars were first discovered by \cite{Web95}.  \\
The discovery of J0133$-$1302 with the KAT-7 telescope was achieved as a result of the large field of view of the instrument and its sensitivity to extended and low surface brightness radio emission. However, poor resolution of the KAT-7 radio telescope, due to its small baselines, and small collecting area meant that it could not resolve the various components of the radio source into distinct sources. 
For a deeper and detailed analysis, there was a need for high resolution observations, which we obtained from the Giant Metrewave Radio Telescope (GMRT; \citealt{Swa91}). For the first time, our GMRT observations have resolved the extended sources in \cite{Col16} into new sources which were not previously observed in the KAT-7 and NVSS structures. 
The current paper thus extends the observational effort to detect the components of the GRG and to study the morphology and spectrum of this source. This GRG could be a powerful source with hints of restarting AGN activity.  
\\
We present the details of the GMRT observations and all the data in Section~\ref{sec:robs}, while the results from the radio observations and the analysis are presented in Section~\ref{sec:results}. The summary and conclusions follow in Section~\ref{sec:disconc}. \\
Throughout the paper a flat, vacuum-dominated Universe with $\Omega_m = 0.32$ and $\Omega_\Lambda = 0.68$ and $ H_0=67.3$ km s$^{-1}$Mpc$^{-1}$ is assumed. 
\section{The Data}
\label{sec:robs}
\subsection{GMRT Observations}

The follow up observations were carried out using the GMRT instrument at two frequency bands: 323 MHz and 608 MHz in January 2016. The project code was 29\_004 (PI: N. Mhlahlo). The observation dates were 04 January (608 MHz), 06 January (323 MHz) and 16 January 2016 (608 MHz), with the latter being make-up observations, as the observations of 04 January were unusable due to bad radio frequency interference (RFI). The total observing time on the source was 4 hours for each observation, at the central frequencies of 306 MHz and 591 MHz, with a total bandwidth of 33 MHz.  \\
Data were recorded using an 8-s integration time with the available frequency band divided into 512 channels. 
The uv-data were reduced using the standard tasks in the Astronomical Image Processing System (AIPS; \citealt{Gre03}) software package. Bad data which resulted from RFI and antenna malfunctions were flagged using AIPS tasks TVFLG, WIPER and CLIP.
To calibrate the data in phase and flux we used the phase calibrator source 0116-208 and flux density calibrator sources 3C147 and 3C48. The flux density calibrator sources were also used to do bandpass calibration. The SPLIT task was used to transfer the calibration to the target source, and the 512 channels were averaged to 16 channels at 323 MHz, and to 4 channels at 608 MHz. 
The reduced data were deconvolved using the CLEAN algorithm, and the images were generated by AIPS task IMAGR.
Several rounds of phase-based self-calibration were performed, and primary beam correction was done on the resulting images using the task FLATN.  \\
Furthermore, the same data were reduced using the Source Peeling and Atmospheric Modeling (SPAM) software package \citep{Int09, Int14}. SPAM, which is an extension to AIPS software package, provides fully-automated data reduction scripts for high-resolution, low-frequency radio interferometric observations of the GMRT data. The software includes direction dependent calibration and imaging, different RFI and bad data mitigation tasks, and modeling. The benefits of using SPAM software on GMRT data is that the data reductions are very efficient, highly reproducible, and give good quality results \citep{Int14}.
All the images were loaded into CASA (Common Astronomy Software Applications; \citealt{McM07}) for further analysis.
The resulting maps are displayed in Fig.~\ref{fig:aips-spam-maps}. The top panels contain the AIPS images, and the SPAM images are shown in the bottom panels. In the bottom-right panel the components of the GRG are labelled, where L1 represents the south-eastern lobe and L2 the north-western lobe; C stands for the radio core of the GRG, and B is possibly a background or foreground source, or one of the multiple hotspots of a complex lobe.  \\
In all the images we see diminished emission between the upper and lower part of the southern lobe. This effect is enhanced in the AIPS images especially at 323 MHz which could indicate some missing flux at this frequency. Our study, as a result, will focus on the SPAM images, which are better, as already mentioned above.
\subsection{Archival Data: GLEAM, NVSS and VLASS}
\label{sec:archival}
Details of the GaLactic and Extra-galactic All-sky MWA Survey (GLEAM), which was carried out with the Murchison Wide-field Array (MWA; Lonsdale et al. 2009; Tingay et al. 2013) are described by \cite{Way15} and \cite{Hur17}. We accessed the data from this survey on the MWA Telescope website \footnote{http://www.mwatelescope.org/GLEAM}. We used a set of 20 images with the bandwidth of 7.7 MHz over a frequency range of 76-231 MHz, which is the frequency range covered by the survey. GLEAM has a mean beam size of $\sim$ 2$^\prime$. In the vicinity of our target source the rms varies between 5 and 15 mJy/beam.  \\
For a proper spectral ageing it is necessary to have a good spectrum at many radio frequencies as possible. The low-frequency GLEAM data are important for proper estimation of alpha injection, and despite the larger beam (see Fig.~\ref{fig:GLEAM-GMRT}) and flux uncertainties, we ensured that we used the GLEAM data in a proper way. 
We have taken extreme care to check if there are any background/foreground sources which are not part of the GRG within the immediate vicinity of the GRG, and if their respective fluxes are convolved in the broad beam of GLEAM.  We found that all the external sources are independent of the GRG's emission.
\\
We also used the GRG data from the NVSS, which has a beam size of $45''$, at 1.4 GHz.    
In addition, obtained VLA Sky Survey\footnote{https://science.nrao.edu/science/surveys/vlass} (VLASS) data which are available to the public. VLASS provides coverage of the entire sky visible to the VLA, above declination -40 deg, at high angular resolution and observing frequency of 2.5 arcsec and 3 GHz, respectively \citep{Lac20}. We obtained a quicklook (QL) image with a rms level of $\sim$0.13 mJy/beam where the GRG core is clearly visible. 
\section{The Analysis And Results}
\label{sec:results}
%
Flux density measurements at 323 MHz and 608 MHz were performed with the AIPS task TVSTAT, which adds up all the pixel values within the selected area to give the total flux density of the selected source. We checked the consistency of the flux scale based on five strong point sources in the field of the GRG. We also checked the source positions using the same five point sources for both GMRT maps. \\
The errors in flux density were determined using the equation
\begin{figure}
\centering
\includegraphics*[width=84mm,height=75mm]{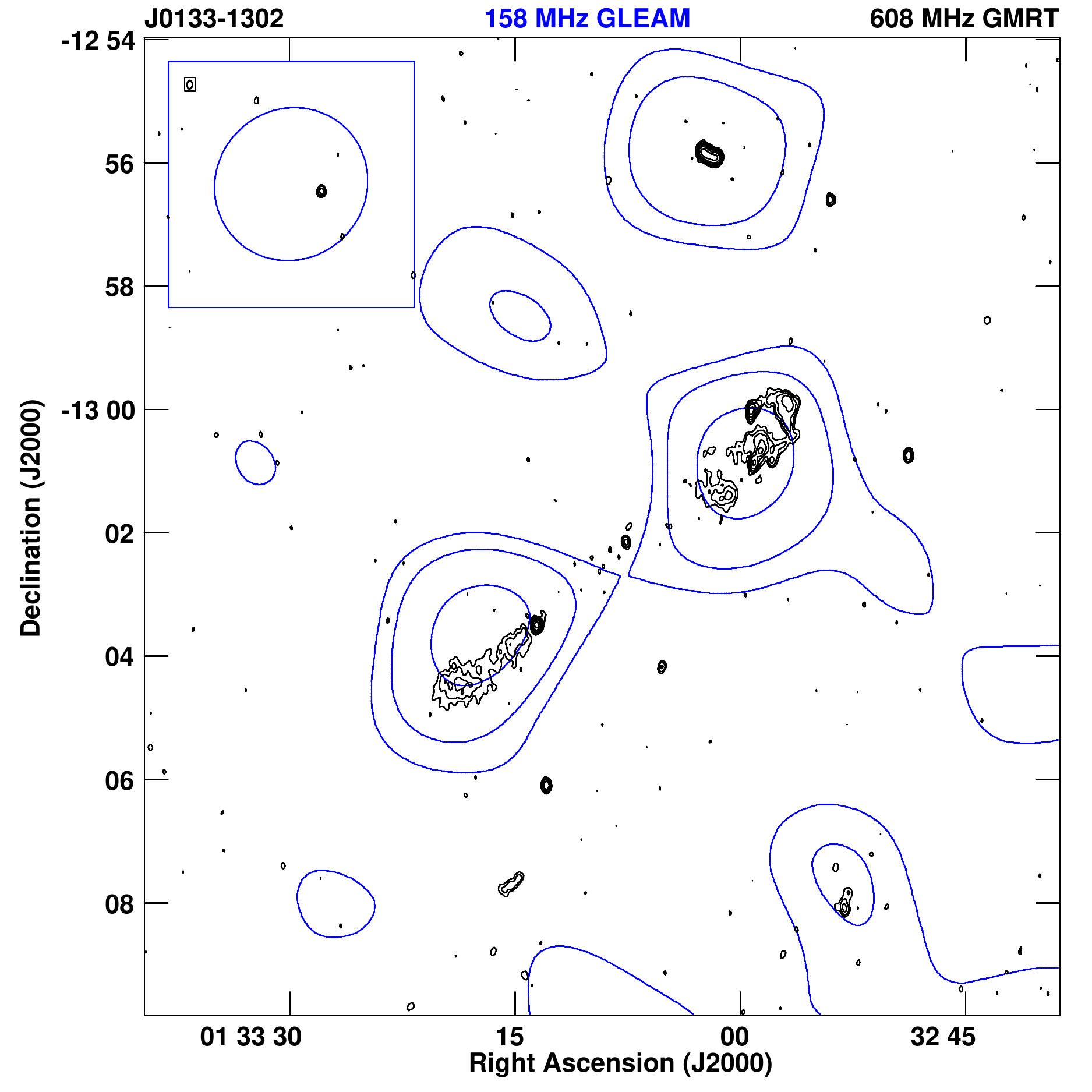}
\caption{Radio image of the GRG J0133$-$1302. Blue contours at 0.03 mJy/beam $\times$ 1, 2, 4 are GLEAM at 158 MHz, and the black contours are from GMRT at 608 MHz with contour levels similar to those in Fig.\,1-bottom right panel. The resulting beams size are shown at the top left hand corner.}
\label{fig:GLEAM-GMRT}
\end{figure}
\begin{table}
\caption{GLEAM flux densities for the lobes of J0133$-$1302 are shown after subtracting the fluxes of the Core and the background B source that were predicted from a quadratic fit and a straight line, respectively (Fig.~\ref{fig_core_bgr_fits}), to the values at higher frequencies. For GLEAM, the calibration errors are approximated at 10\% \citep{Hur17}. The angular resolution of the GLEAM survey is 2.5$^\prime \times 2.2^\prime$ [sec ($\delta+26.7^{\circ}$)] at 154 MHz \citep{Way15}. At the highest frequencies the resolution is $\sim2^\prime$. In addition, the NVSS values are given.}
\label{Tab:flx_archive}
\begin{tabular}{c c c c }
\\
\hline\hline
\multicolumn {1}{c}{Telescope}  & Frequency  &\multicolumn {2}{c}{Flux Density$\pm$ error ($mJy$)}  \\ \cline{3-4}
   &  (MHz) & L1 & L2    \\
\hline
  GLEAM   & 84        &  254.15$\pm$64.62    &  314.92$\pm$68.98    \\  
          & 92        &  232.19$\pm$55.48    &  277.60$\pm$59.24    \\  
          & 99        &  208.29$\pm$54.20    &  278.35$\pm$59.05    \\  
          & 107       &  257.44$\pm$45.57    &  246.50$\pm$46.68    \\ 
          & 115       &  178.13$\pm$37.10    &  209.69$\pm$40.39    \\
          & 122       &  190.66$\pm$35.69    &  235.22$\pm$40.00    \\
          & 130       &  165.46$\pm$32.19    &  196.36$\pm$35.56    \\
          & 143       &  179.83$\pm$35.38    &  241.93$\pm$40.79    \\ 
          & 151       &  107.61$\pm$30.31    &  210.20$\pm$37.18    \\
          & 158       &  156.68$\pm$30.39    &  158.37$\pm$32.82    \\
          & 166       &  127.26$\pm$30.59    &  153.34$\pm$33.61    \\
          & 174       &  164.89$\pm$31.18    &  173.79$\pm$33.69    \\ 
          & 181       &  170.26$\pm$29.84    &  135.33$\pm$29.80    \\
          & 189       &  109.90$\pm$26.17    &  149.89$\pm$30.00    \\
          & 197       &  121.68$\pm$27.32    &  237.68$\pm$35.76    \\
          & 204       &  103.89$\pm$22.38    &  197.48$\pm$29.67    \\ 
          & 212       &  94.43 $\pm$21.77    &  180.04$\pm$27.97    \\
          & 220       &  101.55$\pm$20.87    &  134.07$\pm$23.99    \\
          & 227       &  108.80$\pm$21.86    &  113.90$\pm$23.52    \\
  NVSS    & 1400      &  16.3$\pm$1.5        &  42.98$\pm$2.33      \\
\hline
\end{tabular}
\end{table}
\begin{equation}
  \Delta S=\sqrt{(S \times 0.05)^2 + \Big ( rms \times \sqrt{\frac{area}{beam}} \Big )^2}
\label{eqn: S-errors}
\end{equation}
We calculated the error in the integrated flux densities ($S$) from two error contributions. The first is from an absolute calibration uncertainty, which is estimated at approximately 5\% (similar as \citealt{Ains16, Lal05, Lal07, Mau13}). The second is the rms noise of the map which is taken into account by the second term in eq. \ref{eqn: S-errors}, with $area$ standing for the area of the source and $beam$ being the beam size.
\\
The flux densities at other frequencies were obtained from the archival data (see Section~\ref{sec:archival}). In the GLEAM and NVSS images of the GRG, the GRG core (C) is embedded in the diffuse plasma of the southern lobe, L1. Furthermore, there is what appears to be a background/foreground source (B), or one of the multiple hotspots of a complex lobe near the outer edge of the northern lobe L2, which is unresolved in NVSS and GLEAM images - and there is no redshift or spectral information about B in the literature. As a result, the flux densities of B, C, L1, L2 could not be measured directly from the GLEAM and NVSS maps.
The flux densities of B and C were estimated using the best-fitting power law models at higher frequencies, and extrapolated to the low GLEAM frequencies for which we could not directly measure them due to poor resolution or contamination of the diffuse emission of the lobes.   \\
We first measured the flux densities of sources C and B from the SPAM (323, 608 MHz) and VLASS (3000 MHz) maps where they are resolved or separated, and fitted a line/curve through those data to determine the spectral index (see Fig.~\ref{fig_core_bgr_fits}). We then used the spectral index to predict the core and background source fluxes at the lower GLEAM frequencies and in NVSS.
We then obtained the total flux density values of the entire lobes in GLEAM and NVSS from Vizier\footnote{http://vizier.u-strasbg.fr/viz-bin/VizieR}, from which we subtracted the predicted core fluxes and the background source fluxes. This resulted in the estimated lobe fluxes (which exclude the core and background fluxes) in the GLEAM and NVSS maps which are displayed in Table~\ref{Tab:flx_archive}. For GLEAM, the point at 76 MHz was removed, as the beam is very large and the whole source (northern and southern lobe) is just a single blob.
\begin{table}
\caption{GMRT flux densities and spectral index for the different components of J0133$-$1302 at 325 MHz and 608 MHz.}
\label{Tab:flx-spec_gmrt}
\begin{tabular}{cccc}
\hline\hline
Structure & \multicolumn{2}{c}{Flux Density (mJy)} & Spectral Index   \\
         & 323 (MHz) & 608 (MHz) &  $\pm$ error    \\ 
\hline
 C       &   6.75$\pm$0.42     &  12.99$\pm$0.70    & 0.72$\pm$0.13     \\ 
 L1      &   71.56$\pm$4.09     &  39.67$\pm$2.83   & -0.92$\pm$0.14     \\ 
 L2      &   131.24$\pm$7.38    &  76.23$\pm$4.45   &  -0.79$\pm$0.13   \\  
 S3      &    27.05$\pm$1.51    &  12.57$\pm$0.72    &  -0.92$\pm$0.13    \\  
 B       &   12.57$\pm$0.74     &  9.70$\pm$0.56     &   -0.45$\pm$0.13   \\  
 S4a     &   51.76$\pm$2.68     &  29.40$\pm$1.52    &   -0.79$\pm$0.12    \\  
 S4b     &   34.84$\pm$1.84      &   23.33$\pm$1.22    &   -0.60$\pm$0.12  \\    S5      &   3.51$\pm$0.41      &   1.57$\pm$0.21    &   -0.92$\pm$0.28     \\
\hline
\end{tabular}
\end{table}
The errors in the GLEAM flux density were estimated as (e.g. \citealt{Whit17})
\begin{equation}
  \Delta S=\sqrt{(S \times 0.1)^2 + rms^2}
\label{eqn: S-errors-gleam}
\end{equation}
with the calibration uncertainty being 10\%, and the total error $\lesssim 25\%$.
\begin{figure}
\vspace{0.8cm}
\hbox{
\includegraphics[scale= 0.5,width=84mm,height=75mm]{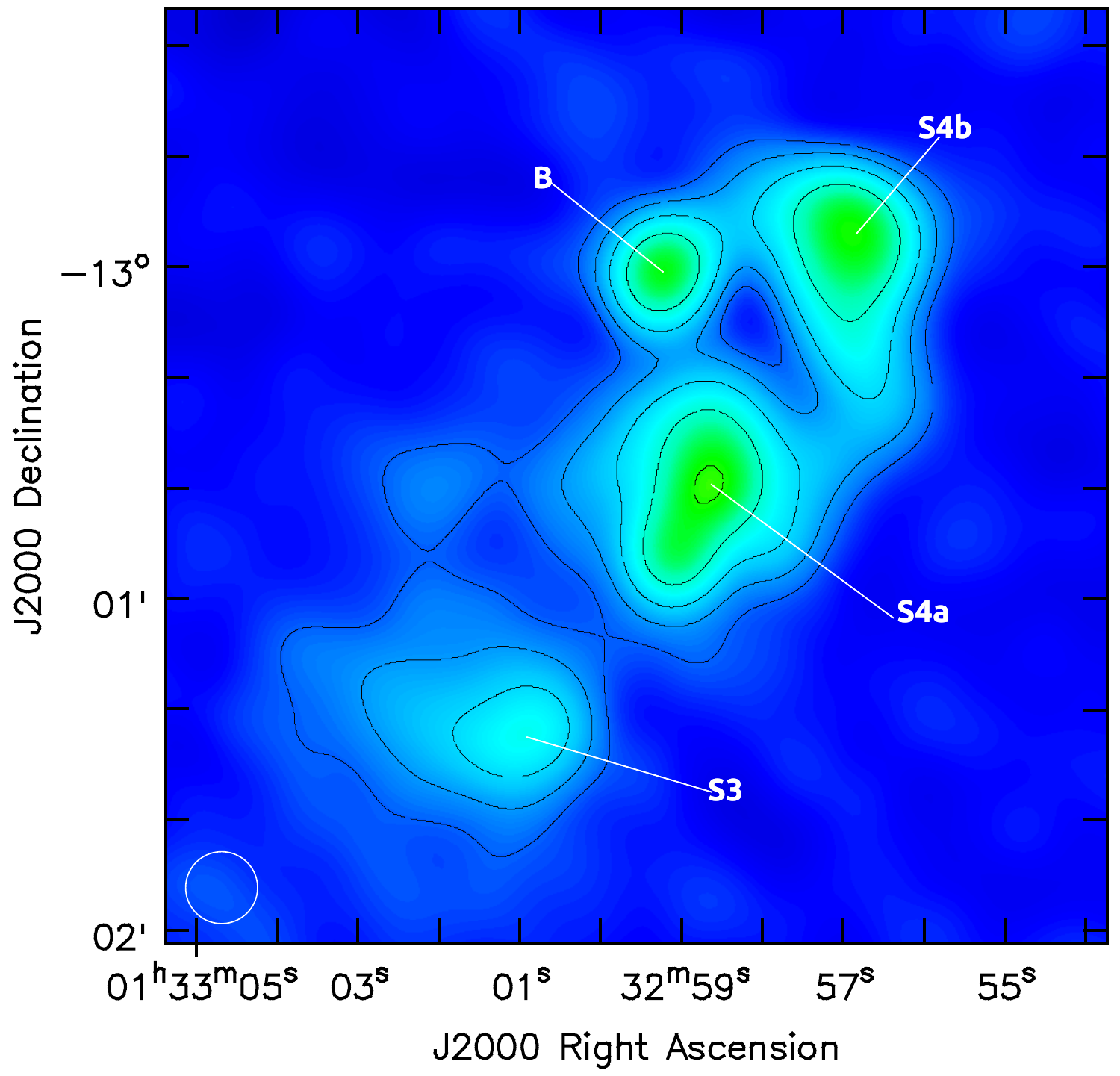}
}
\caption{A close-up look at the northern lobe in Fig.~\ref{fig:aips-spam-maps}  (bottom right panel).}
\label{fig:L2_zoom}
\end{figure}
The total errors for the predicted core and background source fluxes are also estimated at $\lesssim 25\%$.  \\
The fluxes of all the GRG components in the GMRT data are displayed in Table~\ref{Tab:flx-spec_gmrt}.
These fluxes, which include those of the new observed radio sources B, S3, S4a and S4b (see Fig.~\ref{fig:L2_zoom}) which are not resolved in the KAT-7 and NVSS images, were measured directly from the SPAM maps at 323 MHz and 608 MHz.  
In addition to these fluxes, the flux density of the core in NVSS at 1400 MHz was measured to be 18.8$\pm$3.0 mJy. The background source is not resolved in the NVSS map and its flux could not be measured. Furthermore, since the VLASS map shows only the core and the structure B, the flux density was measured only for the core and the background source. The flux density measurement of the core at 3000 MHz gave 21.85$\pm$1.16 mJy, and that of the background source was found to be 4.87$\pm$0.51 mJy.   \\
\subsection{Source Morphology}
\begin{figure*}
\centering
\includegraphics*[width=150mm,height=125mm]{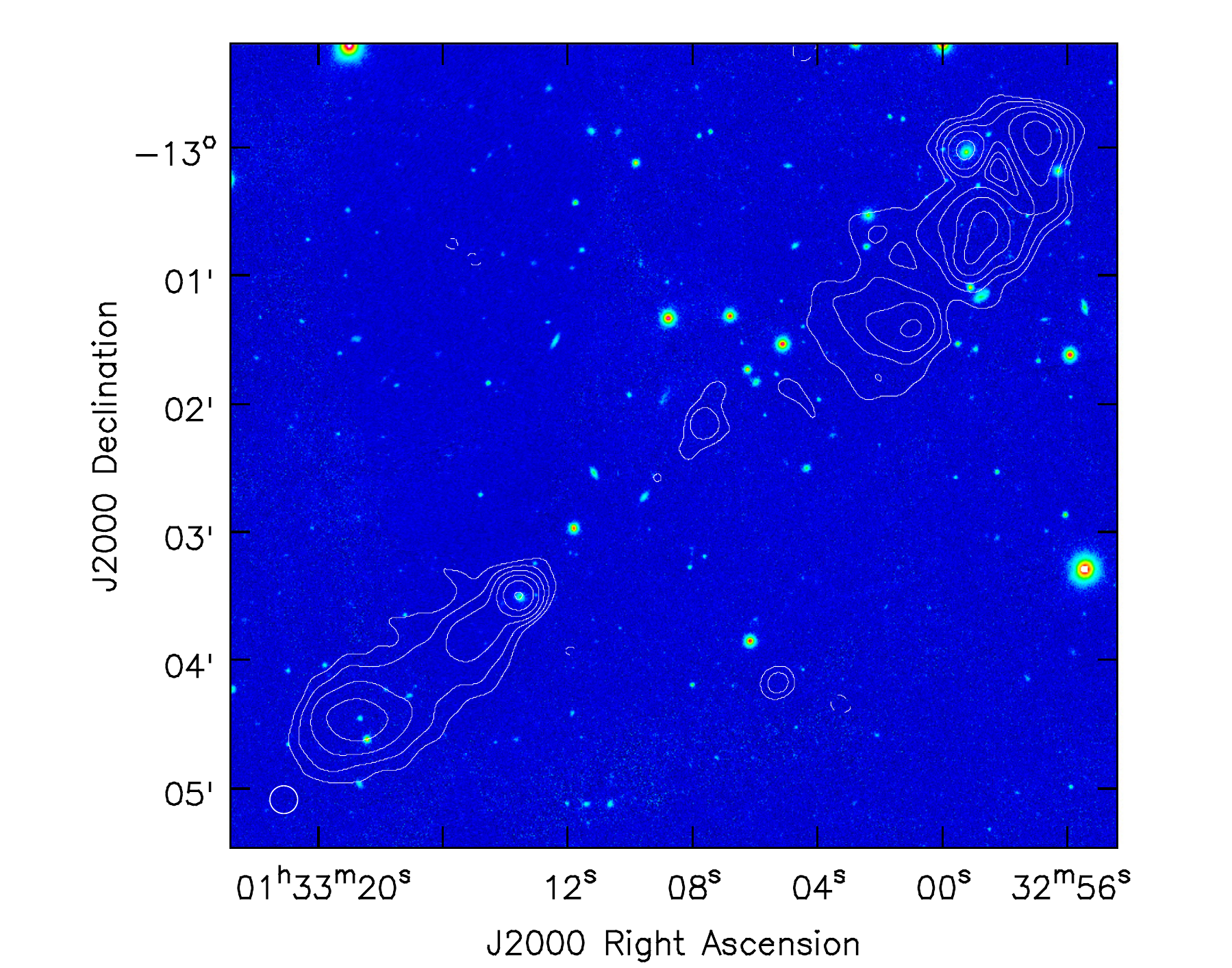}
\caption{Radio contours of the smoothed GRG image at 608 MHz are overlaid on the smoothed optical red $^{\prime}$band$^{\prime}$ image from PanSTARRS. The restoring beam of the image is 8$^{\prime\prime}$ $\times$ 5$^{\prime\prime}$ (PA=-2 deg) and is shown in the bottom-left hand corner. Radio contours start at 0.3 mJy/beam ($3\times rms$) and then scale by a factor of 2.}
\label{fig:spam-nvss_conto_dss2}
\end{figure*}
\begin{figure*}
\centering
\includegraphics*[scale=0.5,width=170mm,height=120mm, trim = {0 1.9cm 0 3.0cm}, clip]{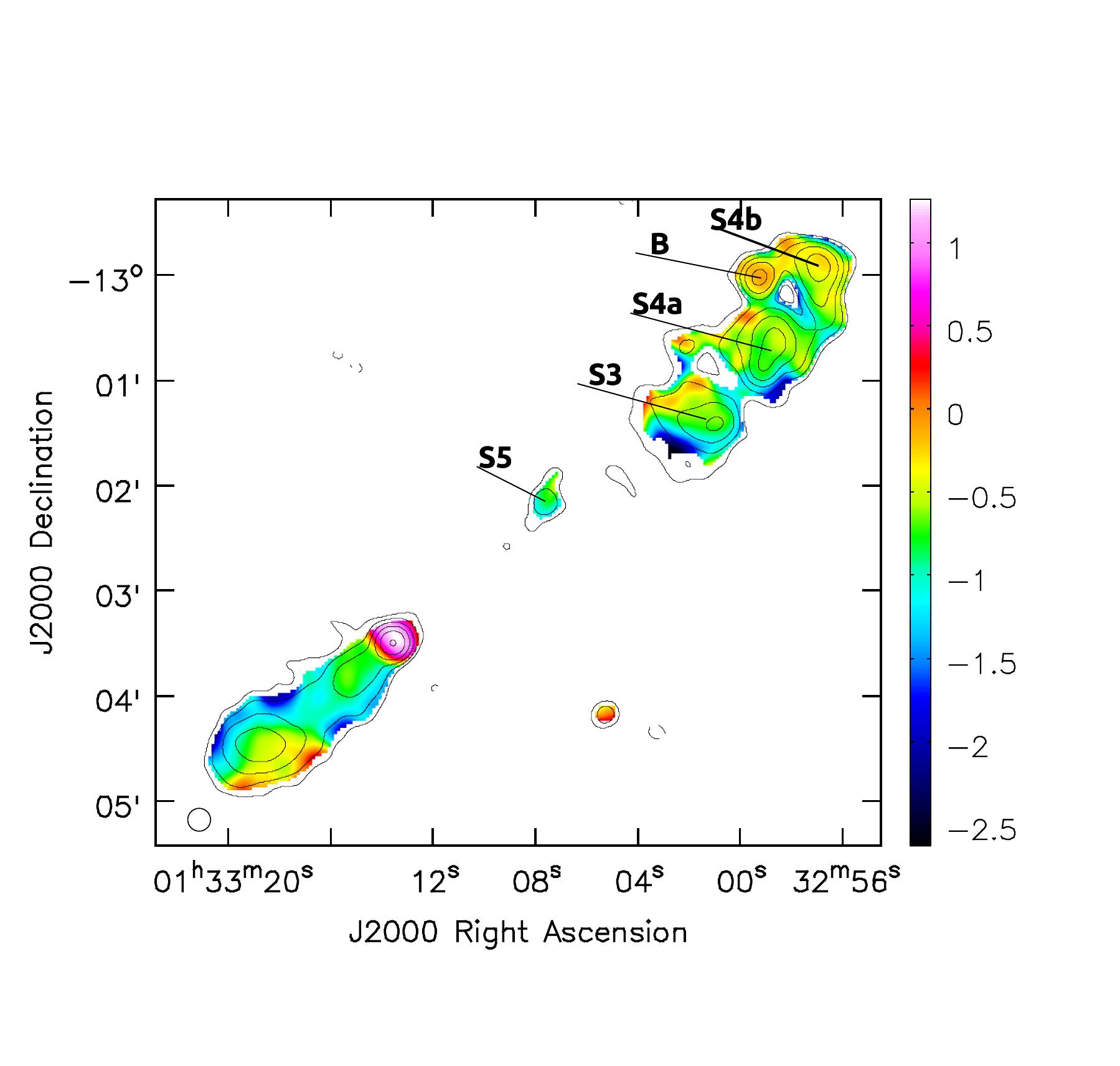}
\caption{Spectral index image of J0133$-$1302.  Radio contours of the GRG in GMRT at 608 MHz are overlayed on the colour-scale image of spectral index between 323 MHz and 608 MHz. The contours are at -3, 3, 6, 12, 24, 48, 96 mJy/beam times the noise level. The spectral index scale is shown at the right hand side of the figure.}
\label{fig:spec-indx-map}
\end{figure*}
\begin{figure}
\vspace{0.2cm}
\centering
\includegraphics*[width=90mm, height=120mm]{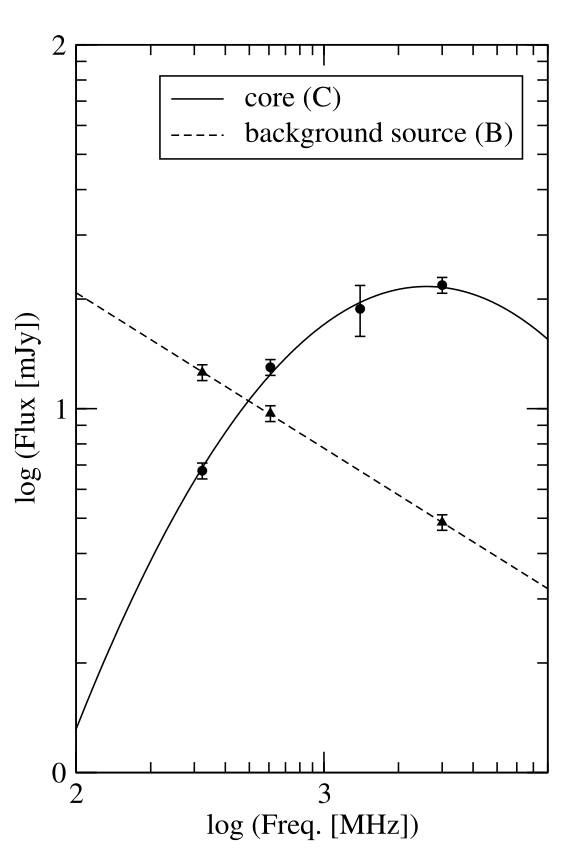}
\centering
\caption{The spectra of the core, and of the background source B. The exact frequencies of the data points shown in this plot are 323, 608, 1400, 3000 MHz in the case of the core and 323, 608, 3000 MHz in the case of the background source. The exact flux density values and errors of these data points are given in Table\ref{Tab:flx-spec_gmrt} and in the text.}
\label{fig_core_bgr_fits}
\end{figure}
\begin{figure*}
\centering
\includegraphics*[width=75mm, height=60mm]{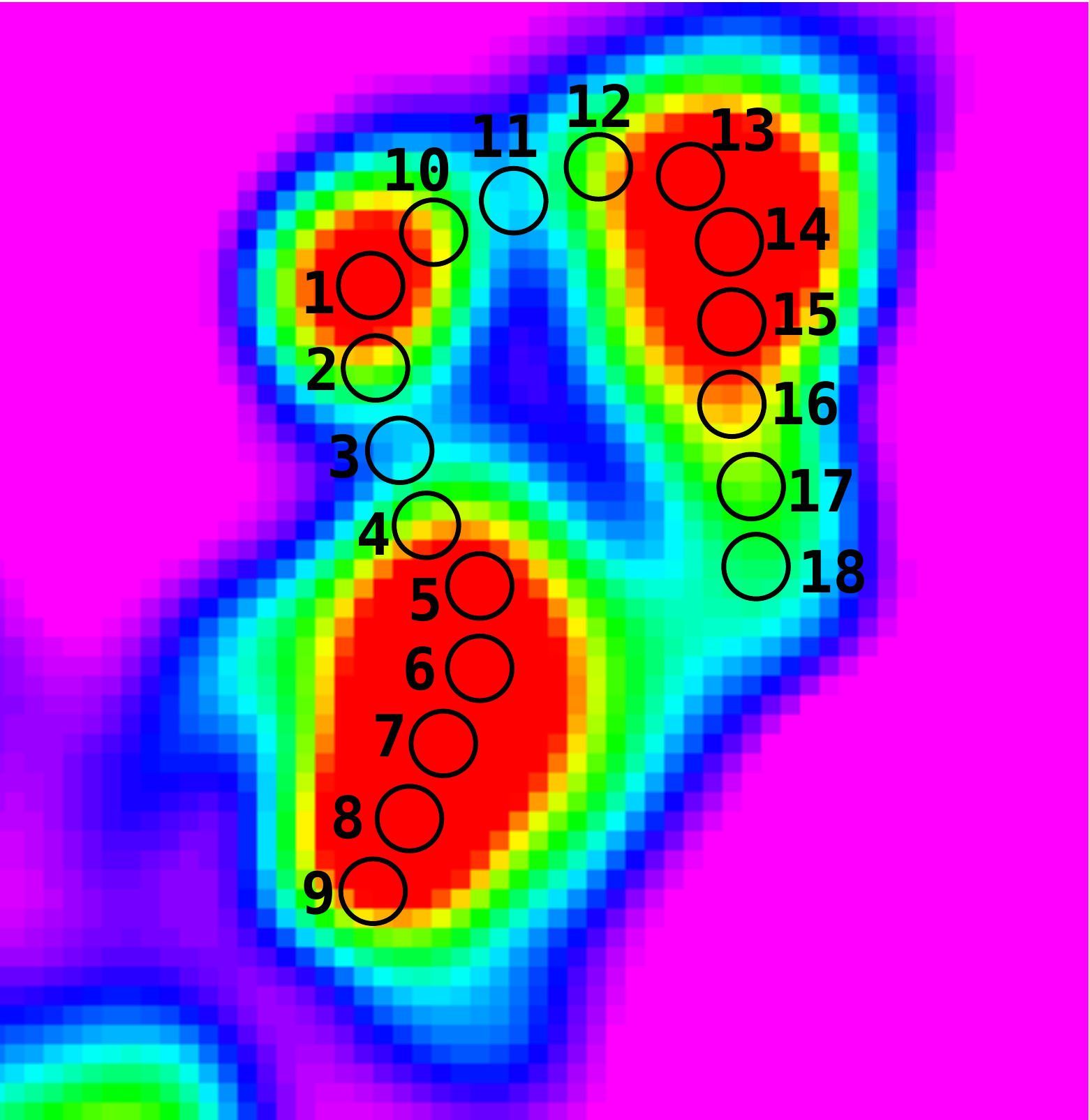} \\[\smallskipamount]
\includegraphics[scale=0.4]{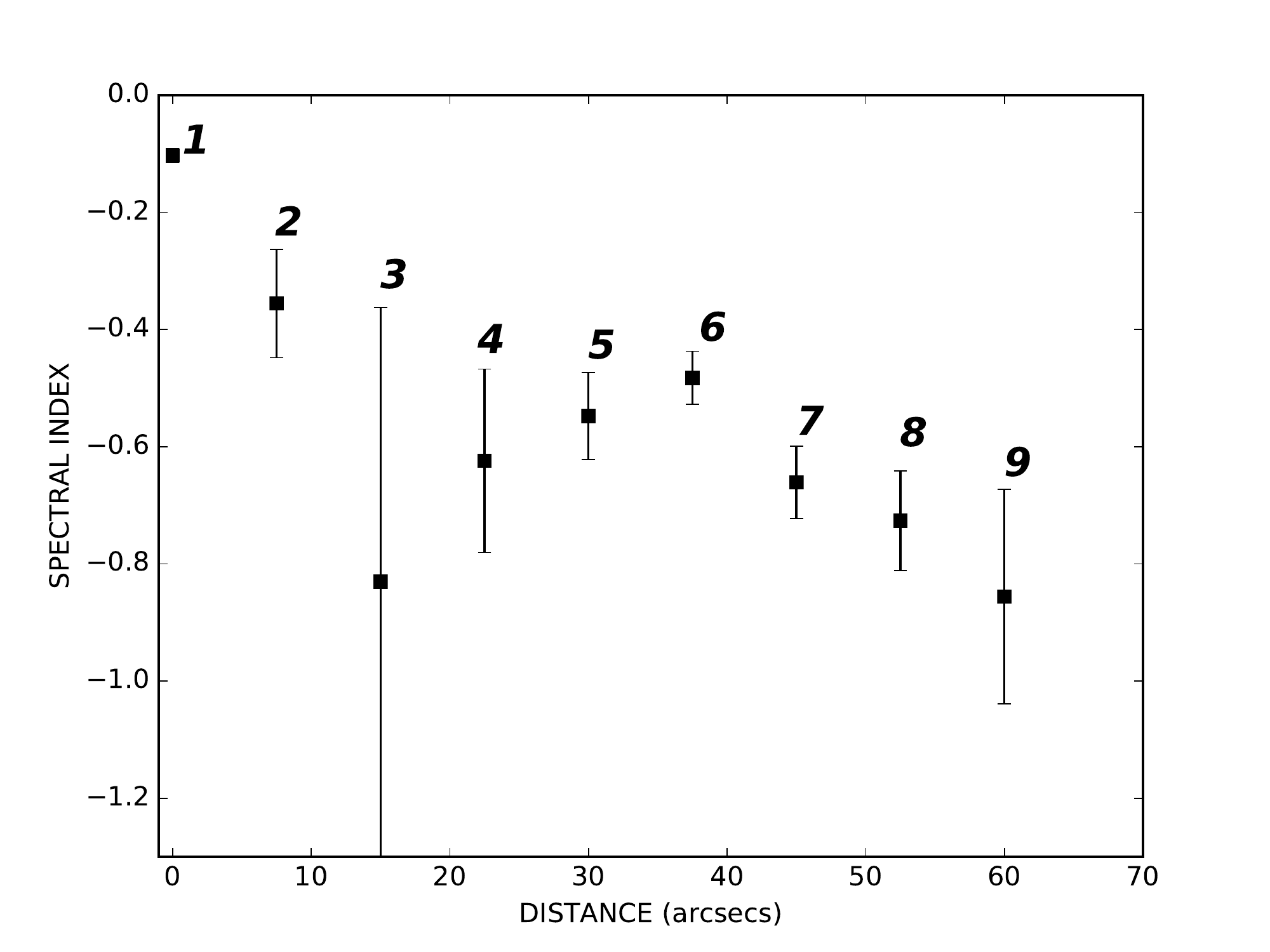} 
\includegraphics[scale=0.4]{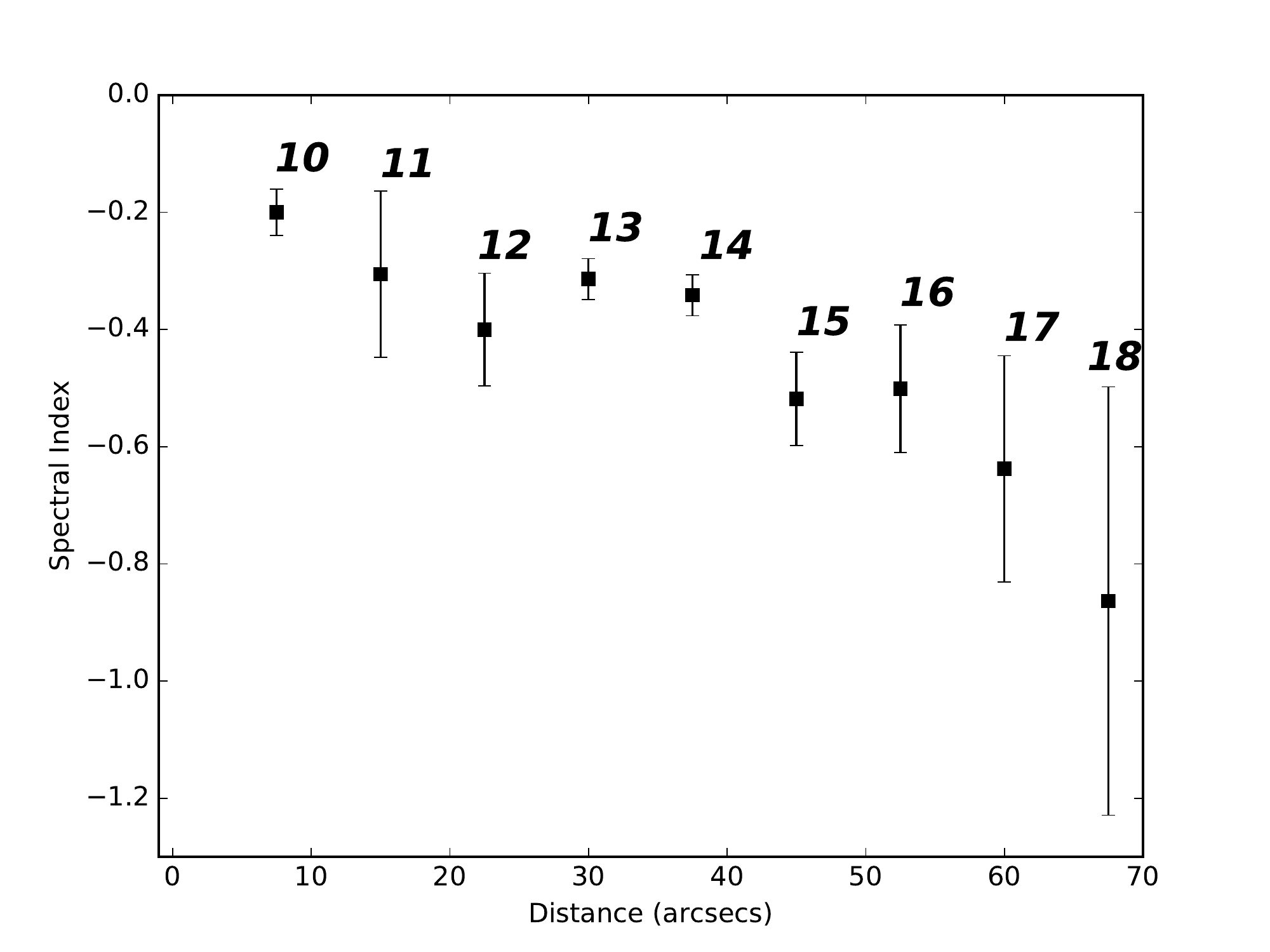}
\hspace{2cm}
\includegraphics[scale=0.4]{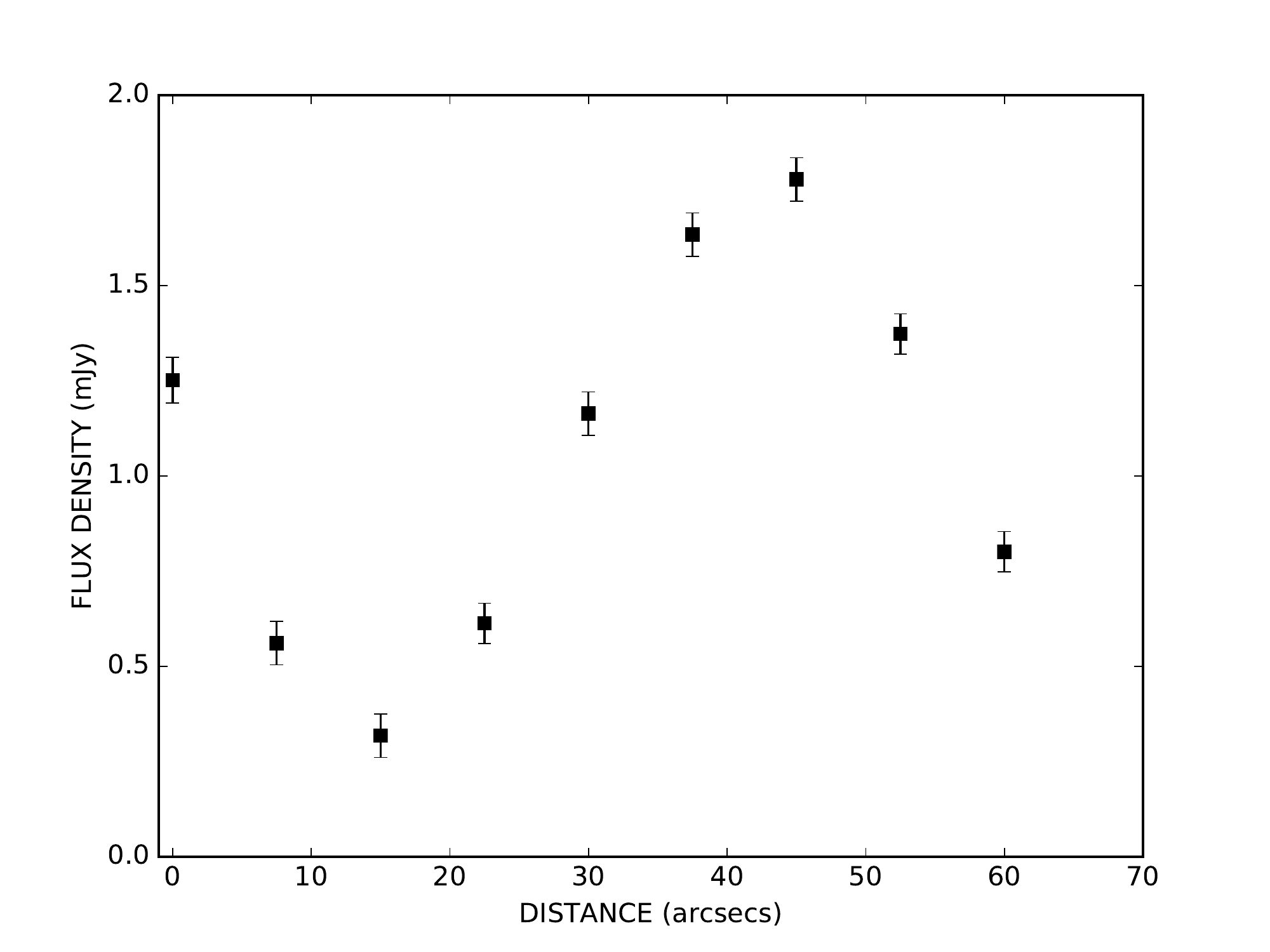}  
\includegraphics[scale=0.4]{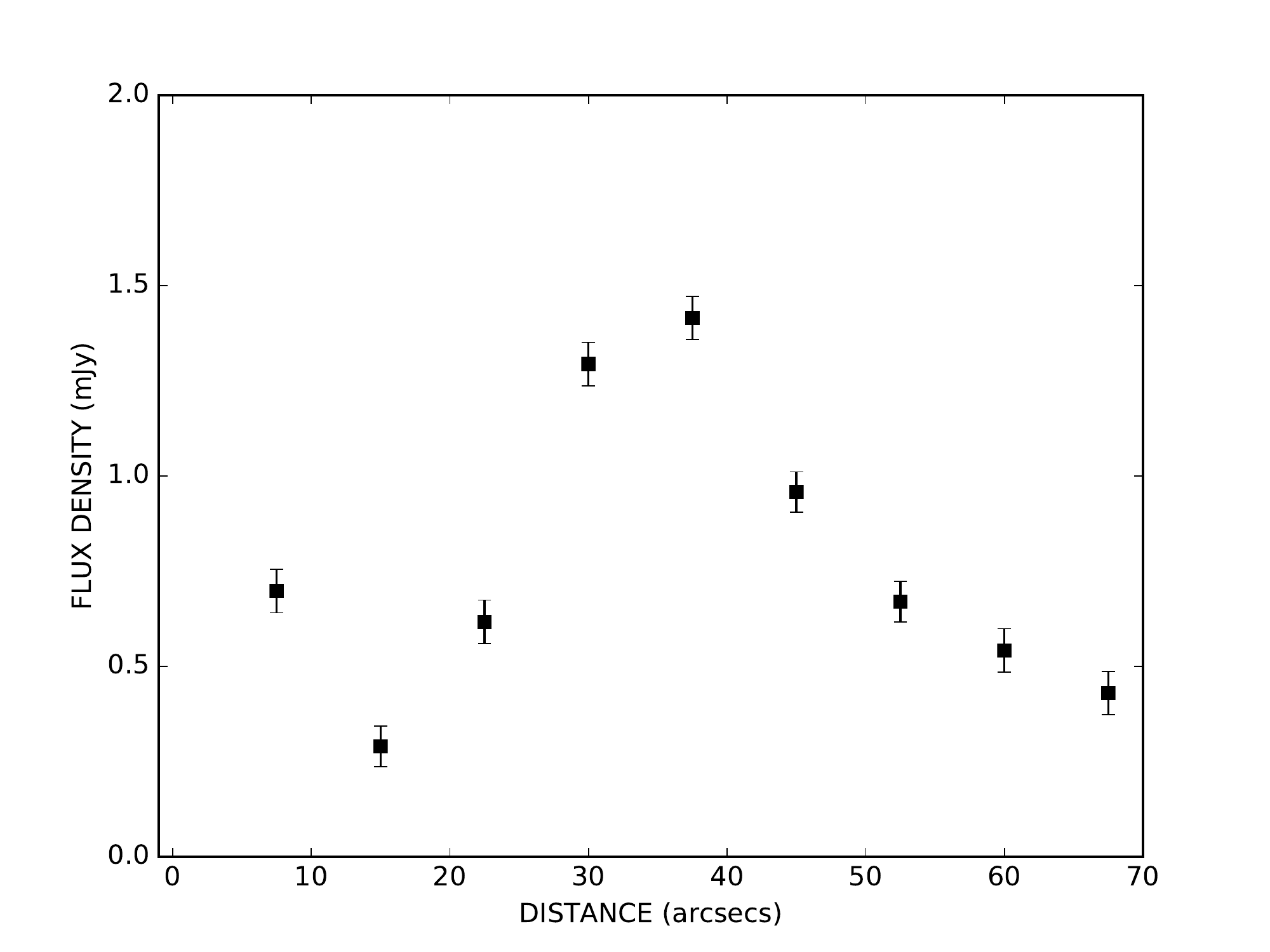}
\caption{Top panel: Regions where spectral analysis was performed are indicated by the circles, along the ridge lines of the proposed DBT.  Middle panels: showing spectral distribution along the circles labeled 1-9 (left) and 10-18 (right). Bottom panels: corresponding flux profiles at 323 MHz along the same circles labeled 1-9 (left) and 10-18 (right) following the same numbering order. Profiles at 608 MHz show similar trends. }
\label{fig:DBT-analysis}
\end{figure*}
\begin{figure*}
\centering
\includegraphics[width=90mm]{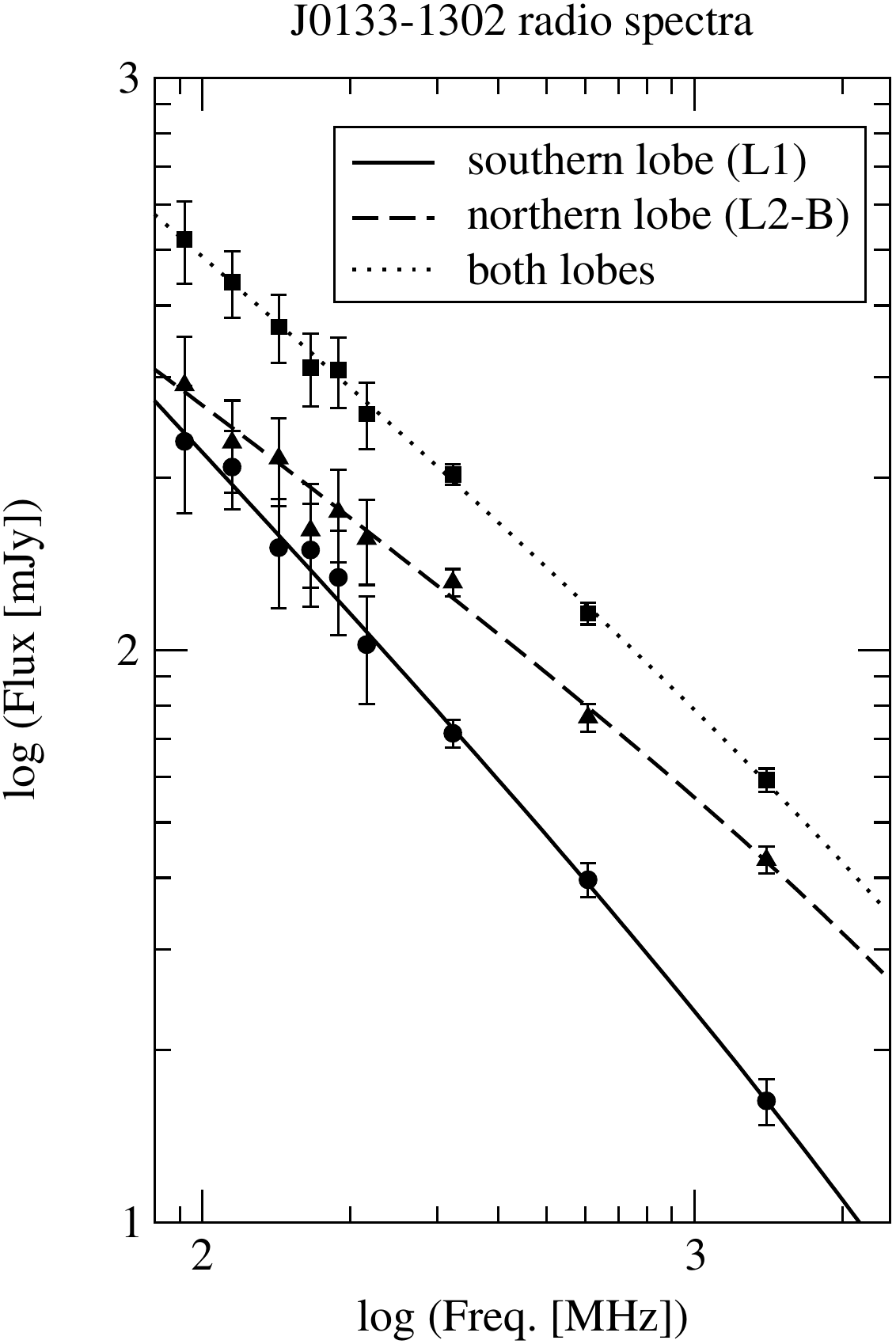}
\caption{The total spectrum of the lobes of J0133$-$1302. Each GLEAM point is an average of three GLEAM fluxes. The curves represent the spectral ageing JP model fit. The exact frequencies of the data points shown in this plot are
92, 115, 143, 166, 189, 215.75, 322.88, 607.96, 1400 MHz.}
\label{fig:model_plots}
\end{figure*}
As it can be observed in Fig.~\ref{fig:aips-spam-maps}, J0133$-$1302 has a clear core detection at the two GMRT frequencies, and an optical counterpart (Fig.~\ref{fig:spam-nvss_conto_dss2})\footnote{https://panstarrs.stsci.edu}. The core appears less bright at lower radio frequencies (Table~\ref{Tab:flx-spec_gmrt}), and gets brighter as the frequency increases, with the highest brightness recorded at 3000 MHz.
All the radio images in Fig.~\ref{fig:aips-spam-maps} seem to show a double-lobed structure, albeit with an axially asymmetric orientation. Lobe L1 shows a clear connection to the radio core C, with a slight decrement in flux at the waist of the lobe (which is more enhanced in the AIPS image at 323 MHz). This apparent connection of the southern lobe with the core indicates that a jet might be present. The angular size of L1 is $\sim 2^{\prime}$, corresponding to a projected linear size of $\sim$ 554 kpc. In constrast, lobe L2 presents a complicated structure, and does not appear to be connected to the core.
For the first time, our GMRT observations have resolved source S4 in \cite{Col16} into three sources, which are S4a, S4b, and B as shown in the zoomed image in Fig.~\ref{fig:L2_zoom}. The complex structure of L2 presents a number of scenarios pertaining to the morphology of J0133$-$1302.  \\
In Fig.~\ref{fig:spam-nvss_conto_dss2}, source B can be identified with an optical source and could be a background/foreground source, while S4a and S4b, together with S3, could be part of the GRG's northern lobe emission. In this scenario, the angular size of L2 would be $\sim 5.4^{\prime}$ ($\sim 1496$ kpc). Alternatively, B could be the core of a DBT radio galaxy, with S4a and S4b being its lobes. In the latter case, only S3 would form the GRG northern lobe emission, with the angular size of L2 would be $\sim 3.7^{\prime}$ ($\sim 1024$ kpc), and that of the entire source being $\sim 6.3^{\prime}$ ($\sim 1745$ kpc). \\
Another possibility is that source B could be one of the hotspots of a complex lobe. \cite{Bes99} showed that nine out of eleven of their sources selected from the 6C sample of Eales displayed multiple hotspots in one or both lobes, where two of the sources, 0825+34 and 1011+36, which contain double hotspots, have an inverted spectrum radio core. \cite{Har07} also observed more than one radio hotspot in at least one of the lobes of a small sample of nearby classical double radio galaxies from which they also detected X-ray emission. Two hotspots have also been observed in the radio galaxy Pictor A \citep{Per97} where it was postulated that the presence of its hotspots suggests that this object probably was recently re-activated.  In this scenario, just as in the first, all the northern sources would constitute L2, and, considering B and S4b have the possible hotspots of the northern lobe, the size of the entire structure for both scenarios would be $\sim 7.8^{\prime}$ (2161 kpc).  \\
\cite{Col16} found the angular size for the entire northern lobe to be $\sim4^{\prime}$ and that for the southern lobe to be $\sim1.2^{\prime}$, corresponding to projected linear sizes of $\sim$ 1107 kpc and $\sim$332 kpc, respectively.
The value obtained by \cite{Col16} agrees with the value we obtained for the scenario where S3 is regarded as the northern lobe hotspot or the northernmost extent of the GRG. However, our value of the angular size for L1 is larger than theirs by slightly less than a factor of two. This is due to the fact that most of the southern lobe emission was convolved with the core emission in \cite{Col16}.
In the scenario where all the northern sources (except B) are part of a giant structure, only S4b would be the hotspot. In that case, we would expect the flux density of S4b to be higher than that of S4a, however, the flux density of S4a is higher than that of S4b at both GMRT frequencies (Table~\ref{Tab:flx-spec_gmrt}).  \\
The conservative approach would be to regard S3 as the northern lobe hotspot or the northernmost extent of the source. In this case we would expect S3 to show a similar spectrum to that of L1. 
Under the assumption that B is the core of a DBT RG (meaning that only S3 is part of the GRG northern lobe as already mentioned above), and that the lobes of the putative DBT RG are symmetric in flux, the higher flux density of S4a would indicate the mixing of the emission both from the DBT RG's lower lobe (S4a) and the GRG's upper lobe (associated with S3) at or near the location of S4a. This suspected mixing of emission should lead to spectral steepening at S4a.  \\
We do not observe any compact and pronounced hotspots in L1.  \\
We note that J0133$-$1302 presents a highly asymmetric large-scale structure where the core is shifted towards the southern lobe. There are two main effects that have been used to explain structural asymmetry in radio galaxies: relativistic beaming (e.g. \citealt{Ahl19}) and jet-environment interactions, via jet entraintment (e.g. \citealt{Per14}), jet stalling \citep{Mas16} or failed collimation of an initially conical jet by the environment \citep{Kra12} as well as lobe-environment interactions which are important at large scales and which are supported by dynamical radio source models (e.g. \citealt{Har18}) and numerical simulations \citep{Har13,Har14}. There are sources, however, whose asymmetric structure cannot be understood easily by ascribing it either to orientation and relativistic beaming effects, or to different kinds of interaction with the environment or asymmetric distribution of gas in the environment
\citep{Sai96}. Some of these sources are characterised by the apparent absence of a hotspot either on one lobe (e.g. J1211$+$743: \citealt{Pir11}), or both lobes (e.g. 0500$+$630 (4C63.07): \citealt{Sai96}, J1918$+$742: \citealt{Pir11}). In the former case, a diffuse radio component on one side of the galaxy is observed while the opposite component appears edge-brightened with a prominent hotspot.
This also includes Hybrid morphology radio sources (HyMoRS) which are a rare type of radio galaxies that display different FR types on opposite sides of their nuclei \citep{Gop00, Gaw06, Kap17}.
For some of these sources the radio jet is facing the diffuse lobe (e.g. J1211$+$743: \citealt{Pir11}). J0133$-$1302, to some extent, seems to show a structure that is similar to the two sources that were studied by \cite{Pir11} as it shows a diffuse southern lobe without a prominent hotspot at the outer edge, and a northern lobe with a possibility of the presence of a single or multiple hotspots. \\ However, recently \cite{Rod19} investigated the relationship between asymmetries in radio source properties and those of the environment where they quantified the asymmetry in the radio source environment through optical galaxy clustering which they argue provides a proxy for the ambient gas density distribution which interacts with the radio lobes. They found that the length of radio lobes in FRII sources is anticorrelated with both galaxy clustering and lobe luminosity, suggesting that the environment is the cause of radio source asymmetry. More light on the cause of the asymmetry of J0133$-$1302 will be shed by future studies that will involve quantifying the environment of J0133$-$1302, that will also be useful for testing the predictions of the orientation unification model (\citealt{Mul08}, \citealt{Spi21}) for this source.  \\
To probe the entire structure of J0133$-$1302 we used the symmetry parameters of this source, and observed that it is very asymmetric in its arm-length (separation) as well as flux density ratios. We have determined the arm-length ratio (AR) and the flux ratio (FR) for two scenarios: the one where S4b has the possible hotspot of the northern lobe (AR, FR = 3.35, 1.8, respectively), and where S3 is the northern lobe hotspot, or the northernmost extent of the source (AR, FR = 2.4, 2.6, respectively).  
%
%
\cite{Ish99} found the median value of the higher luminosity GRGs in their sample to be about 1.39 for the arm-length ratio, while \cite{Kon08}, for their sample of 10 giant radio sources, found a median value of $\sim$1.31, and both values are smaller than those we have obtained above, pointing to the exceptionally large asymmetry of our source. Also, the flux ratios for our source are larger than theirs, except for one source (if it is compared to scenario 1 above).\\
%
%
%
Since the southern lobe has no pronounced hotspot, we cannot constrain the beaming effect in J0133$-$1302. 

\subsection{Spectral Index Analysis}
\subsubsection{GRG lobes and core}
The spectral index distribution of the lobes of a GRG can provide crucial information about its history. 
The spectral index map of J0133$-$1302 was obtained by comparing the SPAM images at 323 MHz and 608 MHz. The resolutions of the two images were 12.27$^{\prime\prime}$ $\times$ 6.70$^{\prime\prime}$ at 323 MHz and 7.94$^{\prime\prime}$ $\times$ 5.09$^{\prime\prime}$ at 608 MHz. To produce the spectral index map with the same beam and cell size the following steps were taken: i) the images were convolved using the task IMSMOOTH in CASA (since the resolutions of the two images were very close), and the beams were made to be circular, with a resulting beam size of 13$^{\prime\prime}$ $\times$ 13$^{\prime\prime}$, ii) the geometry of the maps was aligned 
using the task HGEOM in AIPS, iii) the task COMB was then used to produce the spectral index image. \\
The spectral index map is shown in Fig.~\ref{fig:spec-indx-map}, revealing a very flat radio core spectrum, with a mean spectral index ($S_{\nu} \propto \nu^{-\alpha}$) of $\alpha^{0.6}_{0.3} = 0.72 \pm 0.13$. The spectrum in the lobes is patchy, with regions of very steep and relatively flatter spectrum. The mean spectral index in the lower lobe is $\alpha^{0.6}_{0.3} = -0.92 \pm 0.14$ (exluding the two flatter tiny patches in the south-west side of the lobe) while it is $\alpha^{0.6}_{0.3} = -0.79 \pm 0.13$ for L2 (after excluding only source B).  \\
In Fig.~\ref{fig_core_bgr_fits} are the spectra for the radio core and the background source. The observed $S-\nu$ relation was fitted with  
 \[ \log \left [\frac{S}{\mathrm{mJy}} \right] = -5.75+4.15 \log \left [ \frac{\nu}{\mathrm{MHz}} \right ] - 0.61  \left( \log \left [\frac{\nu}{\mathrm{MHz}} \right ] \right)^2 \]
%
for the core, and with
%
\[ \log {\left[\frac{S}{\mathrm{mJy}} \right]} = 2.17-0.43 \log \left[ \frac{\nu}{\mathrm{MHz}} \right] \]
%
for the background source. The core exhibits an inverted spectrum, and there is an expectation for the fit to go down at higher frequencies. 
The flat and inverted spectrum of the core indicates the presence of energised relativistic electrons, and points to dynamical activity of a recurrent activity radio galaxy \citep{Sai09}. Such sources are characterised by their flat spectral index of the core, and relic/dying lobes. The spectrum of the resurrecting central engine becomes much flatter than those of the diffuse emission, which is what we observe for J0133$-$1302.
We note that the spectral index of S3 ($\alpha^{0.6}_{0.3} =-0.92\pm 0.13$) is very similar to that of L1 ($\alpha^{0.6}_{0.3} =-0.92\pm 0.14$), a strong indication that among the northern sources, only S3 belongs to J0133$-$1302. The smaller structure, S5 (see Fig.~\ref{fig:spec-indx-map}), could be part of L2. Its spectral index value ($\alpha^{0.6}_{0.3} =-0.91\pm 0.23$) is very similar to that of S3 and L1. The measured flux and spectral index values for all the GRG components are tabulated in Table~\ref{Tab:flx-spec_gmrt}.
\subsubsection{Putative DBT radio galaxy core and lobes}
As already indicated, sources S4a ($\alpha^{0.6}_{0.3} =-0.79\pm 0.12$) and S4b ($\alpha^{0.6}_{0.3} =-0.60\pm 0.12$) could be connected to the compact source B ($\alpha^{0.6}_{0.3} =-0.45\pm 0.13$) (see Fig.~\ref{fig:spec-indx-map}).  
To investigate this possibility, in Fig.~\ref{fig:DBT-analysis} we show the spectral index distribution of the proposed DBT source. 
We computed the spectral index at nine different positions for both the southern blob (S4a) and the northern blob (S4b) at 323 MHz and 608 MHz, starting at B, and at about 7.5$^{\prime\prime}$ from B, respectively, and going towards the end of the tail, averaging the spectral index values within small circular bins (3$^{\prime\prime}\times3^{\prime\prime}$ in size).
The circles, along the ridge lines of the proposed DBT, defining the regions where we did our analysis, are shown in Fig.~\ref{fig:DBT-analysis} (top panel).  \\
If the proposed source were a typical tailed radio galaxy, it would be very flat at, and near the nucleus. And as one gets into the brighter parts of the tails, the spectral index might get a bit flatter again. But then, as one progressed further along, there should be clear steepening, as one went off the nucleus. 
This is what we observe (middle panels). The spectral index is flatter in the regions closer to the nucleus for S4a ($\alpha^{0.6}_{0.3} =-0.1 \div -0.35$) and S4b ($\alpha =-0.2 \div -0.3$) where more energetic electrons are continuously injected by the central engine, and then it steepens along the lobes, as one moves futher from B, reaching values of ($\alpha \sim -0.8 \div -0.9$) for both lobes. 
The flattening in the southern blob, from $20^{\prime\prime}-40^{\prime\prime}$, and at 20$^{\prime\prime}$ in the northern blob, should indicate where the flux is going up. This can be seen in the bottom panels of Fig.~\ref{fig:DBT-analysis}, which show the variation in flux densities as one moves away from B. 
This suggests that the spectral behaviour, of both the northern and southern parts of the proposed source, is consistent with what is expected from a tailed RG.
The steepening of the spectral index along the tail of the DBT RG implies that the relativistic electrons that are responsible for the radio emission suffered energy losses after their first ejection from the nucleus B. Furthermore, the average spectral index of S4a is steeper than that of S4b. This asymmetry in spectral index could be the result of mixing of emission between S3 and S4a, and might be an indication of some interaction between the two sources. 
Investigating such interaction would require redshift information on this proposed DBT source, as well as data at more than two frequencies for a detailed spectral index analysis.
\subsection{Source Energetics}
The spectrum of J0133-1302 was initially fit with the JP \citep{JP73} and continuous injection (CI: \citealt{Pac70}) models of radiative losses which describe the time-evolution of the emission spectrum from particles with an initial power-law energy distribution characterised by the injection spectral index $\alpha_{inj}$ and distributed isotropically in pitch angle relative to the magnetic field direction.  \\
We found that the CI model does not provide a good fit to the data (e.g. has worse values of the reduced $\chi^{2}$). In addition, its application to the lobes of J0313$-$1302 is physically unjustifiable. This is because there could hardly be any compact hotspots, and it seems that the jets are no longer feeding the lobes. Therefore, the JP model is more appropriate to fit the spectra of this source. However, caveats related to this model outlined e.g. by \cite{Harw13, Harw15, Harw17} need to be borne in mind. In applying the JP model, our assumptions are: (i) the radiating particles after entering the lobes are not re-accelerated, (ii) the magnetic field lines are completely tangled and the field strength is constant throughout the energy-loss process, (iii) the particles have a constant power-law energy distribution and (iv) the time of isotropization of the pitch angles of the particles is short compared with their radiative lifetime.  \\
Using the SYNAGE software package \citep{Mur96}, the JP model fit to the spectra of the lobes was performed, with the normalisation, the injection spectral index ($\alpha_{inj}$) and the break frequency ($\nu_{br}$) left as free parameters. Fitting results are shown in Fig.~\ref{fig:model_plots}. The break frequency was found to be $\nu_{br} = 22.1$ GHz for the southern lobe and $\nu_{br} = 19.7$ GHz for the northern lobe.  However, one has to remember that the break frequency is based on an extrapolation of the measurement. The fitted injection spectral index, $\alpha_{inj}\sim -0.89$ of the L1 southern lobe is different from the fitted injection spectral index of L2 (all northern sources excluding B) which is $\alpha_{inj}\sim -0.62$. In other words, while the injection spectral index of the lower lobe is in agreement with what we measure from the spectral index image ($\alpha \sim -0.92$), that of the upper lobe does not agree with the image spectral index average value of $\alpha \sim -0.79$. This could be due to the complicated structure of the northern lobe. \\
It is not easy to recognize conclusively what ingredients the northern lobe consists of.
However, and as already indicated, the fact that the S3 structure has a spectral index value of $-0.92\pm0.23$ which is similar to the spectral index value of L1 (Table~\ref{Tab:flx-spec_gmrt}) could suggest that only S3 belongs to the source J0133$-$1302, whereas S4a, S4b and B constitute a distinct source, which we suggest is a DBT RG, or S4a and B are the hotspots of a complex lobe. \\
Because of the problems associated with the complex structure of the northern lobe of J0133$-$1302, we have attempted to estimate the magnetic field strength and synchrotron age for only the southern lobe L1.  \\
The spectral age can be calculated using
\begin{equation}
  \tau_{rad}=50.3 \frac{B_{eq}^{0.5}}{B_{eq}^2 + B_{CMB}^2} (\nu_{br}(1+z))^{-0.5} {\mathrm{Myr}}
\end{equation}
$B_{CMB}=0.318(1+z)^2$ is the magnetic field strength equivalent to the cosmic microwave background radiation at the redshift $z=0.3$ of our target. \\
Using the minimum energy arguments, we calculated the magnetic field following \cite{Lon11}. Our assumptions included the cutoff frequencies of $\nu_{min} = 10$ MHz and $\nu_{max} =100$ GHz, the filling factor of 1, and the pure electron-positron plasma. The volume of the lobe was approximated by assuming a cylindrical shape of 1.5 arcmin in length and a radius of 0.3 arcmin. This led to the value of the magnetic field strength of $B=1.3\pm0.1$ $\mu G$.  
Our small value of the magnetic field strength is comparable to the values estimated for other GRGs (see e.g. \citealt{Kon08, Mac09}).
Our ageing analysis suggests that at the estimated break frequency of this lobe of 22.1 GHz, the corresponding spectral age is about 11 Myr. However, the above break frequency is located outside the range of frequencies (the highest value is 1.4 GHz) where the flux densities are measured. This implies that the age of the southern lobe should be less than about 44 Myr.
\section{Summary and Conclusion}
\label{sec:disconc}
We have performed an analysis of the radio core and lobes of GRG J0133$-$1302 at multi-radio frequencies using data from the GMRT and public surveys. Our comparison of the radio spectral indices of the lobes and the radio core has revealed a steep spectrum of the lobes, which is contrasted by the flat inverted spectrum of the core. From the spectral index map, the lobes have spectral index values of $\alpha \sim -0.92$ (L1), and $\sim-0.79$ (L2), while the spectral index of the core is $\alpha \sim0.7$. This suggests decaying emission of the lobes and restarting core activity for J0133$-$1302 radio galaxy, where the emission is breaking out once more, leading to the spectrum of the central structure being inverted and much flatter than those of the diffuse sources which were created by the previous cycle of AGN activity.  At the redshift of $\sim 0.3$ \citep{Col16} L1 has a linear size of $\sim 0.6$ Mpc, and the entire source has a projected linear size of $\lesssim$2\,Mpc. \\
We found that the structure of the northern lobe is complex, and this has led us to identify four possibilities regarding the overall structure of the GRG which are:  \\
i) all the northern sources (except B) are part of a giant structure L2, where only S4b could be the remnant hotspot, \\
ii) all the northern sources are part of a giant structure L2, where B and S4b could be the remnant hotspots of a complex lobe. \\
iii) only S3 forms the GRG northern lobe, and subsequently \\
iv) the source B is possibly the core of a DBT RG, and S4a and S4b are its lobes. \\
Our spectral index analysis seems to support ii), iii) and iv) for the following reasons:   \\
a) concerning ii), sources B and S4b have flatter spectra when compared to the other northern sources and thus could be hotspots. \\
b) regarding iii), S3 and L1 have very similar spectral index values and are steep, which is an indication of decaying lobe emission of a restarting AGN. This suggests S3 belongs to the GRG. \\
c) and regarding iv), our spectral analysis provides evidence that suggests that both the northern and southern parts of the proposed DBT source show a behavour that is consistent with what is expected from a tailed RG. The proposed DBT shows steepening of the spectral index along the tail, which implies that the relativistic electrons that are responsible for the radio emission suffered energy losses after their first ejection from the nucleus B. \\
%
DBT radio sources are generally associated with clusters of galaxies \citep{Bla00, Deh14} and have been used to find nearby and distant clusters (up to $z\sim2$, e.g. \citealt{Bla03, Deh14}). This association of DBT sources with clusters has been supported by recent simulations (see \citealt{Mgu15}) which have shown that clusters with masses above $10^{13}h^{-1}$ M$_\odot$ typically host at least one DBT radio source at some epoch, and those with masses in excess of $10^{15}h^{-1}$ M$_\odot$ are likely to contain multiple DBT radio sources (subject to AGN duty-cycle and projection effects). However, contrary to these expectations and the simulations by \cite{Mgu15}, a study by \cite{Obri18} using a small sample of DBT sources found that there are a few DBT sources within known clusters, and that most of the DBT sources do not reside in clusters. They note that while the former observation may be explained by sensitivity and resolution effects, the latter casts doubt on the current models of bent-tail galaxies. \\ 
We find that the radial distance from the ACO209 cluster center to the background source B is 8 Mpc, which is much larger than the size of this cluster which is at a redshift of $\sim$0.206. This suggests our DBT source lies outside the cluster environment, 
and could be one of the DBT galaxies that should be found outside massive clusters \citep{Mgu15, Obri18}.\\
It is known that the presence of a tailed RG can heavily influence some of the giant radio sources, e.g. radio halos, by supplying relativistic electrons to those sources \citep{Gio93}. In our case, S4a could be associated with the emission in the northern part of the northern lobe of the GRG. In other words, the tailed RG's southern lobe (S4a) might be embedded in, or passing through, the northern part of the upper lobe of the GRG (S3). In this situation the GRG upper lobe might have supplied relativistic electrons to the tailed RG.  \\
Having shown indications of restarting activity in the nucleus, J0133$-$1302 can be added to the list of a few sources where there is episodic activity in an AGN but the source does not have a large-scale DDRG morphology (\citealt{Her17}; see also \citealt{Mah19, Dab20, Dab20a}). 
Thorough multi-frequency analysis combined with environmental analysis is necessary to study the morphological and spectral details of these sources. Furthermore, these radio sources will be perfect objective for the upcoming deep, wide-field MeerKAT and Square Kilometre Array (SKA) surveys that will have the potential to study their population evolution up to high redshifts.

\section*{Acknowledgements}
NM would like to thank Lawrence Rudnick for the insightful discussions on the proposed DBT source.  We thank the anonymous referee whose comments have helped to improve the paper. This work is based on the research supported by the National Research Foundation of South Africa (grant number 111735). MJ was supported by Polish NSC grant UMO-2018/29/B/ST9/01793. We thank the staff of the GMRT, who made these observations possible. GMRT is run by the National Centre for Radio Astrophysics of the Tata Institute of Fundamental Research.
\section*{Data availability}
Data is available on request from the authors.
\bibliographystyle{mnras}

\input{pdraft_GRG_09-2021_arXiv.bbl}

\end{document}